\documentclass[11pt]{article}

\usepackage{hyperref}
\usepackage{a4wide}

\usepackage{mathrsfs}
\usepackage[T1]{fontenc}
\usepackage{mathpazo}
\usepackage{setspace}
\usepackage{amsfonts}
\usepackage{amssymb}
\usepackage{amsmath}
\usepackage{epsfig}
\usepackage{latexsym}
\usepackage{color}
\usepackage{graphicx}
\usepackage{nicefrac}
\usepackage[latin1]{inputenc}
\usepackage{pstricks}
\usepackage{slashed}

\author{
  \begin{minipage}{.97\linewidth}
    \vspace{1cm}
    \begin{center}
      \begin{small}
        \textbf{F. Bourliot}\footnote{bourliot@cpht.polytechnique.fr} ${\ }^1$,
        \textbf{J. Estes}\footnote{estes@cpht.polytechnique.fr} ${\ }^1$,
        \textbf{P.M. Petropoulos}\footnote{marios@cpht.polytechnique.fr} ${\ }^1$ and
         \textbf{Ph. Spindel}\footnote{philippe.spindel@umh.ac.be} ${\ }^2$
      \end{small}
    \end{center}
    \vspace{0.5cm}
    \hspace{2cm}\begin{minipage}{.7\linewidth}
     {\it \begin{footnotesize}
  \begin{itemize}
              \item[${}^1$] Centre de Physique Th\'eorique, CNRS--UMR 7644,
        Ecole Polytechnique, \\
        91128 Palaiseau Cedex, France\\
               \item[${}^2$] Service de M\'ecanique et Gravitation, Universit\'e de
        Mons--Hainaut, \\
        20 Place du Parc, 7000 Mons, Belgique\\
        \end{itemize}
     \end{footnotesize}}
    \end{minipage}
    \vspace{0.5cm}
  \end{minipage}
}

\date{\today}

\title{\vspace{2cm}
 \boldmath \begin{huge}
    \textbf{Gravitational instantons, self-duality and geometric flows}
  \end{huge} \unboldmath
}

\newcommand{\be}{\begin{equation}}
\newcommand{\ee}{\end{equation}}

\newcommand{\bea}{\begin{eqnarray}}
\newcommand{\eea}{\end{eqnarray}}

\begin{document}

\begin{titlepage}
  \maketitle
  \thispagestyle{empty}

  \vspace{-11.5cm}
  \begin{flushright}
    CPHT-RR038.0409
  \end{flushright}

  \vspace{10cm}

  \begin{center}
    \textsc{Abstract}\\
  \end{center}
We discuss four-dimensional "spatially homogeneous" gravitational instantons. These are self-dual solutions of Euclidean vacuum Einstein's equations with potentially non-vanishing cosmological constant. They are endowed with a product structure $\mathbb{R} \times\mathcal{M}_3$ leading to a natural foliation into three-dimensional subspaces evolving in Euclidean time. For a large class of three-dimensional subspaces, the dynamics coincides with the geometric flow on the three-dimensional homogeneous slice, driven by the Ricci tensor plus an $so(3)$ gauge connection.  The metric on the three-dimensional space is related to the vielbein of the three-dimensional subspace, while the gauge field is inherited from the anti-self-dual component of the four-dimensional Levi--Civita connection.

\end{titlepage}

\onehalfspace
\newpage


The aim of the present letter is to report on a relationship between four-dimensional self-dual gravitational instantons and three-dimensional geometric flows. This is a follow-up of former scattered observations about Bianchi IX spatially homogeneous vacuum self-dual solutions of Einstein gravity \cite{Cvetic:2001zx, Sfetsos:2006, Bakas:2006bz, Petropoulos:2008}. Our framework is that of four-dimensional Euclidean geometry $\mathcal{M}_4$ which is topologically  $\mathbb{R}\times\mathcal{M}_3 $. The leaves $\mathcal{M}_3$ of this foliation are assumed to be homogeneous spaces of Bianchi type. The slicing is adapted to the splitting of the action of the SO(4) group in the tangent space into self-dual and anti slef-dual parts in such a way that the anti-self-dual part acts only on the subspace tangent to the homogeneous slicing $\mathcal{M}_3$.

The developments we will exhibit are two-fold. On the one side we show that real, non-degenerate,
self-dual solutions  exist only for unimodular Bianchi groups or the one of type III\footnote{Bianchi III is an exception that will not be discussed here in detail because it lies outside of the geometric-flow correspondence.} and
are classified in terms of the homomorphisms of $\mathfrak{g}\to so(3)$, where $\mathfrak{g}$ is the real Lie algebra of the Bianchi group under consideration and $SO(3)$ the anti-self-dual factor of the group $SO(4)$ acting on the orthonormal vierbein. On the other side, we observe that the self-duality requirement leads to first-order equations, which turn out to be geometric-flow equations for a family of three-dimensional Bianchi manifolds, driven by the three-dimensional Ricci tensor combined with a flat $SO(3)$ gauge connection in the following manner (tildes refer to the three-dimensional quantities as opposed to their four-dimensional counterparts):
\begin{equation}
\frac{\mathrm{d}\tilde g_{ij}}{\mathsf{d}t}=-\tilde R_{ij} -\frac{1}{2}  \mathsf{tr} \left( \tilde A_i  \tilde A_j\right).
\end{equation}

Our motivations for this analysis can be summarized as follows. On the one hand, gravitational instantons are important ingredients of general relativity, both as classical solutions and potentially as tools to handle quantum transitions. Despite many results and solutions of Einstein's equations in the above simplified framework (see e.g.  \cite{Taub-nut, Eguchi:1978xp,Eguchi:1978gw, Belinsky:1978ue, Gibbons:1979xn, Lorenz:1983} -- the list is not exhaustive), no unified pattern is available that captures all Bianchi classes in a simple and comprehensive way.

On the other hand geometric flows of three-dimensional homogeneous spaces are
interesting in their own right and turned out to play a role in Hamilton's program  for proving Poincar\'e's and Thurston's \cite{Thurston:1982} conjectures. From a physicist's perspective, a relevant question is to ask whether and how this flow behavior of one-parameter families of three-dimensional spaces is related to the Euclidean-time evolution inside a gravitational instanton, where the homogeneous spaces appear as the leaves of the foliation. This question is motivated by several facts.

Firstly,  Ricci-flow equations are equivalent to renormalization-group  equations for two-dimensional sigma models with  $t\propto -\log \mu$ \cite{AlvarezGaume:1981hn,Braaten:1985is,  Friedan:1980jm,Osborn:1989bu}. Setting a relation between this renormalization-group  time and the Euclidean time of a gravitational instanton would be one more indication in favor of the dynamical generation of time in string theory -- similar in spirit to the role of the Liouville field in non-critical string theory.
Secondly, we refer to the  recent attempt to modify the ultraviolet behavior of gravity
\cite{Horava:2008ih, Horava:2009uw} by assuming a  foliation of the four-dimensional spacetime  with a privileged time direction, \emph{at the level of the action}, which has drastic consequences for the number of propagating degrees of freedom \cite{Charmousis:2009tc}. There, the further detailed balance condition effectively sets a dynamics where time evolution is a geometric flow on the leaves of the foliation. Last, one should keep in mind that the appearance of first-order equations in gravitational settings is reminiscent of holographic situations, and could ultimately be useful in reconstructing the bulk fields by flowing the boundary data. In the context at hand, this statement could be made more precise following \cite{Mansi:2008br,Mansi:2008bs}.

\begin{table}
\centering
\begin{tabular}{|c|c|l|}
  \hline
  Type & Group & Structure constants \\ \hline \hline
  $\mathrm{I}$        & translations & $c^i {}_{jk} = 0$   \\ \hline
  $\mathrm{II}$        & Heisenberg & $c^1 {}_{23} = -c^1 {}_{32} = +1$  \\ \hline
  $\mathrm{VI}_{-1}$ & $E(1,1)$ & $c^2 {}_{13} = -c^2 {}_{31} = -1$, \, $c^3 {}_{12} = -c^3 {}_{21} = -1$  \\ \hline
  $\mathrm{VII}_0$    & $E(2)$ & $c^2 {}_{13} = -c^2 {}_{31} = -1$, \, $c^1 {}_{23} = -c^1 {}_{32} = +1$  \\ \hline
  $\mathrm{VIII}$       & $SL(2,\mathbb{R})$ & $c^1 {}_{23} = -c^1 {}_{32} = -1$, \, $c^2 {}_{31} = -c^2 {}_{13} = +1$, \, $c^3 {}_{12} = -c^3 {}_{21} = +1$ \\ \hline
  $\mathrm{IX}$         & $SU(2)$ & $c^1 {}_{23} = -c^1 {}_{32} = +1$, \, $c^2 {}_{31} = -c^2 {}_{13} = +1$, \, $c^3 {}_{12} = -c^3 {}_{21} = +1$ \\
  \hline
\end{tabular}
\caption{Unimodular Bianchi groups ($c^i  {}_{jk}$ not explicitly given are taken to be zero).}
  \label{tablebianchi}
\end{table}

In this note we present results without all the proofs in detail. A more elaborate discussion will be delayed to a future communication, where the
extension to vacuum solutions with cosmological constant will also be investigated.

As already stated, we seek for Euclidean four-dimensional spaces of the type
$\mathcal{M}_4 = \mathcal{ M}_3\times \mathbb{R}$ with homogeneous spatial sections $\mathcal{ M}_3$. The latter are assumed to be of Bianchi type: a three-dimensional group $G$ acts simply transitively on the leaves, which are therefore locally endowed with the structure of a group manifold  (hence we exclude $H_3, H_2\times S^1$ or $S^2\times S^1$) with  three independent Killing vectors and left-invariant Maurer--Cartan forms  $\{\sigma^i, i=1,2,3\}$ obeying
\begin{equation}
\label{conG}
\mathrm{d}\sigma^i = \frac{1}{2} c^i_{\hphantom{i}jk}\sigma^j \wedge \sigma^k.
\end{equation}
The structure constants can be put in the form (see e.g.  \cite{Ryan:1975jw})
\begin{equation}
\label{str}
c^k_{\hphantom{k}ij} = \epsilon_{ij\ell}n^{\ell k}+\delta^k_ja_i-\delta^k_ia_j,
\end{equation}
from which we read off their trace: $c^j_{\hphantom{j}ij}=2a_i$. Unimodular groups have zero trace and are referred to as Bianchi A. Our choice for the structure constants\footnote{This choice is the one of  \cite{Ryan:1975jw},  except for Bianchi VI$_{-1}$. In our conventions, the matrix $n$ is diagonal, which implies for Bianchi class A algebras that  the structure constants have the following property: $c^{i} _{\hphantom{i}jk}=0$ whenever $i=j$ or $i=k$.}
of this class is presented in Table \ref{tablebianchi}.

The metric for $\mathcal{M}_4$ is in general of the form:
\begin{equation}
\label{offdiagmet}
 \mathrm{d}s^2=  N^2\mathsf{d}T^2 + g_{ij} \sigma^i  \sigma^j,
 \end{equation}
where $ g_{ij}(T)$ are functions to be determined. It is convenient to introduce an orthonormal frame
$\{\theta^a, a=0,1,2,3\}$
 \begin{equation}
\label{diagmet}
 \mathrm{d}s^2=   \delta_{ab} \theta^a \theta^b,
 \end{equation}
by setting:
 \begin{equation}
\label{vierbein}
\theta^0= N \mathsf{d}T
, \quad
\theta^{\alpha}= \Theta^{\alpha}_{\hphantom{\alpha}j}\sigma^j \quad \mathrm{with} \quad
 g_{ij}= \delta_{\alpha\beta}\Theta^{\alpha}_{\hphantom{\alpha}i} \Theta^{\beta}_{\hphantom{\beta}j}
 \end{equation}
%
($\alpha,\beta,\dots$ label orthonormal space frames so that $\{a\} = \{0,\alpha\}$, whereas $i,j,\dots$ correspond to our choice of invariant forms as they follow from Table \ref{tablebianchi} and Eqs. (\ref{conG})).  We will make the convenient gauge choice $N=\Theta= \sqrt{\det g_{ij}}$, and often use another ``time'' defined as
 $  \mathsf{d}t=\Theta \mathsf{d}T$.

The metric elements $ g_{ij}(t)$ or, equivalently, the frame components $ \Theta^{\alpha}_{\hphantom{\alpha}j}$ are determined by imposing Einstein's equations. The torsion-less connection one-form $ \omega^a_{\hphantom{a}b}$ is determined by the
Cartan structure equations. Its Riemann curvature two-form will be denoted $\mathcal{R}^a_{\hphantom{a}b}$ and satisfies the usual cyclic identity ($\mathcal{R}^a_{\hphantom{a}b} \wedge
    \theta^b=0$).  In four dimensions
we can introduce the dual curvature form:
\begin{equation}
\label{dualR}
\bar{\mathcal{R}}^a_{\hphantom{a}b} = \frac{1}{2}
    \epsilon^{a\hphantom{bc}d}_{\hphantom{a}bc}\mathcal{R}^c_{\hphantom{c}d}
    \equiv\frac{1}{2}
    \bar{R}^a_{\hphantom{a}bcd} \theta^c\wedge \theta^d.
\end{equation}
in terms of which Einstein's vacuum equations read:
\begin{equation}
\label{einsvac}
\mathcal{\bar{R}}^c_{\hphantom{c}d} \wedge
    \theta^d=0.
\end{equation}
%
%
Since we are interested in Euclidean solutions, we can impose (anti-)self-duality: $\mathcal{R}^a_{\hphantom{a}b}=\pm\bar{\mathcal{R}}^a_{\hphantom{a}b} $. This is a sufficient (but not necessary) condition to obtain vacuum solutions thanks to the cyclic identity.

For reasons that will become clear in the following we would like to elaborate on the issue of (anti-)self-duality\footnote{More on self-duality can be found in \cite{Atiyah:1978wi} and \cite{Eguchi:1980jx}.}. Spin connection and curvature forms belong to the antisymmetric $\mathbf{6}$ representation of $SO(4)$.  
%
In four dimensions, this group of local frame rotations factorizes as  $SO(3)_{\mathrm{sd}}\otimes SO(3)_{\mathrm{asd}}$ and the connexion $\omega_{ab}$ and curvature $\mathcal{R}_{ab}$ $SO(4)$-valued forms can be reduced with respect to the  $SO(3)_{\mathrm{(a)sd}}$ as $\mathbf{6}= (\mathbf{3}_{\mathrm{sd}},\mathbf{3}_{\mathrm{asd}})$:
\begin{equation}
\Sigma_\alpha=\frac{1}{2}\left(\omega_{0\alpha} + \frac{1}{2} \epsilon_{\alpha\beta\gamma}\omega^{\beta\gamma}\right),\quad
A_\alpha=\frac{1}{2}\left(\omega_{0\alpha} - \frac{1}{2} \epsilon_{\alpha\beta\gamma}\omega^{\beta\gamma}\right)\label{redconn}
\end{equation}
for the connection and
\begin{equation}
\mathcal{S}_\alpha=\frac{1}{2}\left(\mathcal{R}_{0\alpha} + \frac{1}{2} \epsilon_{\alpha\beta\gamma}\mathcal{R}^{\beta\gamma}\right),
\quad \mathcal{A}_\alpha=\frac{1}{2}\left(\mathcal{R}_{0\alpha} - \frac{1}{2} \epsilon_{\alpha\beta\gamma}\mathcal{R}^{\beta\gamma}\right)\label{redcurv}
\end{equation}
for the curvature, which now reads:
\begin{equation}
 \mathcal{S}_\alpha= \mathsf{d} \Sigma_\alpha -\epsilon_{\alpha\beta\gamma} \Sigma^\beta \wedge  \Sigma^\gamma,\quad
 \mathcal{A}_\alpha= \mathsf{d} A_\alpha +\epsilon_{\alpha\beta\gamma} A^\beta \wedge A^\gamma.\label{sdsolcon}
 \end{equation}
It is clear from the above that  $\{\mathcal{S}_\alpha,\Sigma_\alpha \}$ are vectors of $SO(3)_{\mathsf{sd}}$ and singlets of $SO(3)_{\mathsf{asd}}$ and vice-versa for  $\{\mathcal{A}_\alpha,A_\alpha \}$.

It is sufficient to impose that $\mathcal{S}_\alpha$ or similarly $ \mathcal{A}_\alpha$ be zero to solve vacuum Eisntein's equations. For concreteness we will focus on the self-dual solutions, namely those for which
\begin{equation}
\label{sdsol}
 \mathcal{A}_\alpha=0.
\end{equation}
Anti-self-dual solutions are obtained by $O(4)$ parity or time-reversal transformations.

Equations (\ref{sdsol}) are second-order. First-order equations can be obtained by considering the spin connection $A_\alpha $ in Eq. (\ref{sdsolcon}). The simplest solution to  (\ref{sdsol}) is of course
\begin{equation}
\label{sdsolzero}
A_\alpha=0.
\end{equation}
This  first integral raises immediately two questions: (\romannumeral1) is $A_\alpha=0$ leading to consistent self-dual vacuum solutions, and (\romannumeral2) is this unique? Concerning the second question, it is known that
barring global issues, one can always find an  $SO(3)_{\mathrm{asd}}$ local transformation (see e.g. \cite{Eguchi:1978gw}) such that the anti-dual part $A_\alpha$ of the spin connection is set to zero. Although conceptually important, this property leaves open a practical question: since for any self-dual curvature, one can find a frame where the connection is self-dual, one may ask how many different frames are needed in order to exhaust all possible non-equivalent self-dual connections. Put differently, for a given frame, how many non-equivalent non-self-dual connections exist ($A_\alpha\neq 0$), which can be turned to self-dual ones upon appropriate  $SO(3)_{\mathrm{asd}}$ transformation?

Both questions can be answered accurately. Firstly, Eq. (\ref{sdsolzero}) admits non-degenerate\footnote{Non-degenerate means with an everywhere non-vanishing metric determinant. We do not exclude singularities, which do generically appear in gravitational instantons.} solutions for Bianchi A class and Bianchi III only. This can be proven in full generality, but we shall here present the heuristic argument.
We chose for that the metric to be diagonal.  For almost all Bianchi classes this is always consistent and
non-restrictive\footnote{An exception is again Bianchi III, which must be treated separately, without changing the conclusion though.}. This amounts to taking  $ \Theta^{\alpha}_{\hphantom{\alpha}j}
= \delta^{\alpha} {}_{j}\gamma_j$, which leads to (see Eqs. (\ref{diagmet}, \ref{vierbein})):
\begin{equation}
\label{homfol}
\mathrm{d}s^2 =  \mathrm{d}t^2
    +\sum_i \left(\gamma_i\sigma^i\right)^2;
\end{equation}
$\gamma_i (t)$ are now the functions to be determined. Using the Cartan structure equations,
we obtain   $\omega^0_{\hphantom{0}1}=-\frac{\dot \gamma_1}{\gamma_1\gamma_2\gamma_3}\sigma^1$, whereas $\omega^2_{\hphantom{2}3}=\cdots +\frac{\gamma_2}{2\gamma_3}
  c^2_{\hphantom{2}31} \sigma^1+\frac{\gamma_2}{2\gamma_1}
 c^2_{\hphantom{2}32} \sigma^2+\cdots$. The dots stand for other terms which are irrelevant for our purposes. Following (\ref{redconn}), demanding that the anti-self-dual part of the connection $A_1$ be zero, leads to several equations, among which we find the condition that $\gamma_2 c^2_{\hphantom{2}32}=0$. If the algebra is non-unimodular, there are unavoidable coefficients like $c^2_{\hphantom{2}32}$, so that the metric components like  $\gamma_2 $ are thus required to vanish by self-duality. Hence, the metric tensor is not invertible.

Secondly, it is easy to show that there are as many non-equivalent connections $A$ with vanishing anti-self-dual curvature $\mathcal{A}$  as  homomorphisms of $\mathfrak{g}\to so(3)$.
We will refer to them as \emph{branches} of solutions.  It will turn out that in every case there are two such branches.
Qualitatively, the reason is as follows. In general, vanishing anti-self-dual curvature (Eq. (\ref{sdsol})) requires the connection $A$ be a pure $SO(3)$ gauge. Put differently,  $A$ must be of the form $-\mathrm{d}\Lambda \Lambda^{-1}$, where $ \Lambda$ stands for an $SO(3)$ gauge transformation. We know that the geometry of $\mathcal{M}_3$ is locally that of the group $G$ so that each point $\mathbf{x}$ corresponds to a group element $g(\mathbf{x})$. If we consider gauge transformations in $G$ to define $A$, we could in general take $g(\mathbf{x})$ as such a transformation and $A= -\mathrm{d}g g^{-1}$ would just deliver the left-invariant Maurer--Cartan forms i.e. $A_\alpha=\delta_{\alpha i}\frac{\sigma^i}{2}$ (upon appropriate choice of basis and normalization).
But we have in addition that $\Lambda(\mathbf{x}) \in SO(3)$ and as a consequence the resulting pure connection can only be of the form
\begin{equation}
\label{firoreom}
A_\alpha=\delta_{\alpha i}\frac{ \lambda_i}{2}\sigma^i,
\end{equation}
where $\lambda_i$ are each 0 or 1 depending on whether a given generator
of $\mathfrak{g}$ can be mapped to a generator of $so(3)$.
For each homomorphism $\mathfrak{g}\to so(3)$ there is a set of three numbers
$\{\lambda_1,\lambda_2,\lambda_3\}$.
Each of these choices provides an anti-self-dual connection with vanishing curvature.

The above qualitative reasoning can be made precise. We  define general $I_{\alpha i}$ such that  $A_\alpha=\frac{1}{2}I_{\alpha i} \sigma^i$ and introduce this ansatz in (\ref{sdsol}) together with (\ref{offdiagmet}) and (\ref{vierbein}). The equations we obtain are
\begin{equation}
\label{floweqs}
   \dot \Theta_{\alpha i} =  \Theta_{\alpha j}
   \left[  \left(n^{j\ell} -a_k
\epsilon^{kj\ell}
    \right)g_{\ell i} -
     \frac{1}{2} \delta^j_i n^{k\ell }g_{k\ell }
  \right]  - \Theta I_{\alpha i},
\end{equation}
for the components $0i$, plus a constraint on the constants of the motion $I_{\alpha i}$
\begin{equation}
\label{constraints}
I_{\alpha \ell}c^\ell_{\hphantom{\ell}jk}
+ \epsilon_{\alpha\beta\gamma}I^\beta_{\hphantom{\beta}j}
I^\gamma_{\hphantom{\gamma}k}=0,
\end{equation}
for the $ij$ ones, that sets the announced interplay between $\mathfrak{g}$ and $so(3)$. Indeed, by using appropriate transformations, one can bring the $I_{\alpha \ell}$ into a diagonal form with entries $\{\lambda_1,\lambda_2,\lambda_3\}$ taking the values 0 or 1.

To make contact with the existing literature on the search of gravitational instantons in all Bianchi classes it should be  mentioned that Eqs. (\ref{constraints}) lead (in most Bianchi classes) to imaginary solutions. These are actually related to homomorphisms of $\mathfrak{g}$ into real subalgebras of $su(2,\mathbb{C})$, which provide more freedom but do not correspond to genuine instantons. We can summarize as follows the various possibilities, corresponding to the branches of admissible consistent self-duality equations:
\begin{description}
\item[ Bianchi class A] For these, the rank-zero homomorphism that maps $\mathfrak{g}$
to the nul generator of $so(3)$ with $\lambda_i=0$ is always available and leads, as already mentioned, to consistent solutions. Besides the latter, there is always another one (unique up to trivial algebra automorphisms), which is rank-one in types I, II and VII$_0$ where it maps one generator of $G$ onto one of $so(3)$ with a single non-vanishing $\lambda_i$; and rank-three in type IX where it is the isomorphism of $\mathfrak{g}\equiv so(3)$ to itself with all $\lambda_i=1$. The cases of VI$_{-1}$ (algebra $\mathrm{iso}(1,1)$ of $E(1,1)$) and VIII (algebra $sl(2,\mathbb{R})$) are peculiar. Besides the trivial homomorphism, they exhibit respectively a rank-one and a rank-three  homomorphism\footnote{Actually, those algebras possess a boost generator and consequently a eigenvalue in the Cartan--Killing metric of opposite sign.} in $\mathbb{C}$: $\lambda_1=i, \lambda_2= \lambda_3 = 0$ and $\lambda_1=1,\lambda_2=\lambda_3 = -i$. Although the latter are not relevant for real self-dual solutions, they turn out to be necessary in setting the advertised relation with the Ricci flow of three-dimensional Bianchi spaces.
 \item[ Bianchi class B] The rank-zero homomorphism leads in this class to singular metrics. Another rank-one homomorphism exists in all cases but requires systematically a complex mapping (see also \cite{Lorenz:1983}), with the exception of Bianchi III. The latter will be studied elsewhere.
 \end{description}

From now on, we will focus on the Bianchi A class, and turn to  the interpretation of the Euclidean time evolution in the above gravitational instantons as a geometric flow of a family of three-dimensional homogeneous spaces. At the technical level, this interpretation is motivated by the appearance of first order differential equations with respect to Euclidean time, namely by Eqs. (\ref{floweqs}).
For concreteness we would like to carry out first a well studied case, that of Bianchi IX \cite{Gibbons:1979xn}. As in all Bianchi A cases the diagonal metric ansatz is sufficient, we
will for convenience proceed with that choice till the end of the paper,
leaving for the future the general intrinsic analysis. Setting  $ \Theta_{\alpha i}= \gamma_i\delta_{\alpha i}$  leads in general
to Eq. (\ref{homfol}). For Bianchi IX, we consequently take $I_{\alpha i}= (1-\tilde\lambda)\delta_{\alpha i}$, which allows to capture the two cases as: $\tilde\lambda=0$ (isomorphism) or $1$ (trivial homomorphism).
These two cases correspond respectively to the Taub--NUT and the Eguchi--Hanson branches of Bianchi IX. The first-order self-duality equations (\ref{floweqs}) read:
\begin{equation}
\label{ricIX}
  \begin{cases}
      \displaystyle{ 2 \dfrac{\dot\gamma_1}{\gamma_1} = \left(\gamma_2 - \gamma_3
      \right)^2 - \gamma_1^2 } +2\tilde\lambda \gamma_2\gamma_3, \\[1.2em]
  \displaystyle{     2 \dfrac{\dot\gamma_2}{\gamma_2} = \left(\gamma_3 - \gamma_1
      \right)^2 - \gamma_2^2}+2\tilde\lambda \gamma_3\gamma_1, \\[1.2em]
     \displaystyle{  2 \dfrac{\dot\gamma_3}{\gamma_3} = \left(\gamma_1 - \gamma_2
      \right)^2 - \gamma_3^2}  +2\tilde\lambda \gamma_1\gamma_2.
    \end{cases}
\end{equation}
For the Taub--NUT branch ($\tilde\lambda=0$) the observation (already made in \cite{Cvetic:2001zx, Sfetsos:2006, Bakas:2006bz, Petropoulos:2008}) is that Eqs. (\ref{ricIX})
 reproduce the Ricci-flow evolution of a family of three-dimensional Bianchi IX geometries
 \begin{equation}
 \label{3Dgeom}
  \mathsf{d}\tilde s^2 =
   \tilde g_{ij}
     \sigma^i\sigma^j=\delta_{\alpha\beta}\tilde\theta^\alpha \tilde\theta ^\beta,
\end{equation}
which are also of the diagonal type:
  \begin{equation}
 \label{3Dgeomdiag}
 \tilde g_{ij}(t)= \delta_{ij} \gamma_i(t).
\end{equation}

This observation is remarkable and raises many questions that we will try to sort later in the discussion. For the moment we would like to extend this correspondence to all branches and all Bianchi A classes since those are the ones that systematically deliver a  consistent spectrum of self-dual gravitational instantons appearing in two separate branches.

For the Eguchi--Hanson branch ($\tilde \lambda=1$ in (\ref{ricIX})), the family  of three-dimensional Bianchi IX geometries that flow is again given in (\ref{3Dgeom}, \ref{3Dgeomdiag}). Examining the self-duality equations in (\ref{ricIX}), we find on the right hand side terms which reproduce  (through the process described previously) the three-dimensional Ricci tensor, but in this case there is more: an $so(3)$ gauge field $\tilde A$ appears on the flowing three-spheres, which originates from the Levi--Civita anti-self-dual connection $A$. This field reads:
 \begin{equation}
 \label{3DgfIX}
 \tilde A = \tilde A_i  \sigma^i =-\tilde\lambda \delta_{\alpha i}T^\alpha\sigma^i
\end{equation}
where
 $T^\alpha$ are the generators of $so(3)$ in the adjoint, satisfying $ \mathsf{tr} (T^\alpha T^\beta)=-2\delta^{\alpha\beta}$. This $so(3)$ gauge field vanishes for Taub--NUT but is non-zero for Eguchi--Hanson. In both cases, however, its field strength is zero. With this field, Eqs. (\ref{ricIX}) are recast as announced in the beginning:
 \begin{equation}
 \label{basic}
\frac{\mathrm{d}\tilde g_{ij}}{\mathsf{d}t}=-\tilde R_{ij} -\frac{1}{2}  \mathsf{tr} \left( \tilde A_i  \tilde A_j\right).
\end{equation}

The result at hand deserves several comments. The advertised relation, that turns out to be valid for all Bianchi A classes as we will shortly discuss, sets a correspondence between the time evolution in self-dual gravitational instantons foliated with homogeneous leaves and the flow (parametric in time) evolution of homogeneous spaces. For the sake of simplicity, this correspondence has been exhibited here in the case of a diagonal metric, but it holds more generally. The flow equation (\ref{basic}) follows directly from (\ref{floweqs}) with an $so(3)$ gauge field
   \begin{equation}
 \label{3Dgf}
   \tilde A =-\tilde I_{\alpha i} T^\alpha\sigma^i.
  \end{equation}
In order for the correspondence to be valid, the components $\tilde I_{\alpha i} $ are subject
to the constraint
   \begin{equation}
 \label{3Dgfcon}
  \tilde I_{\alpha \ell}c^\ell_{\hphantom{\ell}jk}
+ \epsilon_{\alpha\beta\gamma}\tilde I^\beta_{\hphantom{\beta}j}
\tilde I^\gamma_{\hphantom{\gamma}k}=0,
  \end{equation}
which is nothing but a flatness condition:
   the constraint
   \begin{equation}
 \label{flat}
\tilde F = \mathsf{d}\tilde A +[\tilde A,\tilde A]\equiv 0.
  \end{equation}
The gauge field is a \emph{background field}, it does not flow:
   \begin{equation}
\dot{\tilde A}=0,
  \end{equation}
but contributes to the flow of the metric. The flatness condition has two different solutions: (\romannumeral1)  the : $\tilde A=0$
corresponds to the Taub-NUT branch whereas the  $\tilde A\neq 0$ reproduces the Eguchi--Hanson branch. Of course the flow equation is not gauge-invariant and it was not expected to be since the actual difference between the various branches is a difference of gauge for the anti-self-dual part of the Levi--Civita connection.

The correspondence between self-dual gravitational instantons with Bianchi homogeneous spatial sections, on the one hand, and three-dimensional homogeneous spaces endowed with a background $so(3)$ flat connection (\ref{flat}) and flowing under (\ref{basic}), on the other, holds for all Bianchi A classes.  These are the classes that exhibit several branches of consistent instantons.  Furthermore the correspondence holds for all these branches because the classification principle for the gravitational-instanton branches is the same as the one that classifies the flat $so(3)$ connections over $\mathcal{M}_3$, which is locally $G$: equations (\ref{constraints}) and (\ref{3Dgfcon}) are both flatness conditions, the former for the anti-self-dual part of the Levi--Civita connection, the latter for the $so(3)$ background gauge field. We will again illustrate this correspondence in the case of diagonal metrics for the remaining Bianchi I, II, VI$_{-1}$, VII$_{0}$ and VIII. We consider now more generally (\ref{3Dgeom}) with
\bea
 \tilde g_{ij}(t)= \delta_{ij} \tilde\gamma_i(t).
\eea
We denote the metric coefficients  $\tilde\gamma_i$ since the advertised correspondence  does not assume the three-dimensional part of the four-dimensional metric to be equal to the three-dimensional metric. Similarly, in the diagonal ansatz, we take (\ref{3Dgf}) as an $so(3)$ gauge field with
 \begin{equation}
 \label{3Dgfgendiag}
\tilde I_{\alpha i}=\tilde\lambda_i  \delta_{\alpha i},
\end{equation}
where $\tilde\lambda_i  $ are subject to the constraints (\ref{3Dgfcon}) which now read:
\be \label{CoEqSelfDual}
 \tilde\lambda_i c^{i}_{\hphantom{i}jk}+\epsilon_{ijk} \tilde\lambda_j \tilde\lambda_k
=0
\ee
with no summation on $i,j,k$.
%
%
%
%
Consequently, the geometric-flow equations obtained from (\ref{basic}) can be written as
\bea \label{eqRicdiag}
{\dot {\tilde\gamma}_i \over \tilde\gamma_i} = - \sum_{j,k=1}^3 \frac{1}{4}\left[\left(c^{i} _{\hphantom{i}jk}\right)^2\tilde\gamma_i^2-2 \left(c^{j} _{\hphantom{j}ki}\right)^2 \tilde\gamma_j^2+2c^j _{\hphantom{j}ki}c^k _{\hphantom{k}ij}\tilde\gamma_j\tilde\gamma_k \right] + \tilde\lambda_i^2  \tilde\gamma_j\tilde \gamma_k
\eea
where the dot stands for $\nicefrac{\mathrm{d}}{\mathrm{d}T}=\tilde\gamma_1\tilde\gamma_2\tilde\gamma_3\nicefrac{\mathrm{d}}{\mathrm{d}t}$.
Correspondingly, the self-duality equations ({\ref{floweqs}) read:
\bea \label{eqSelfDual}
{\dot \gamma_i \over \gamma_i}  =  \sum_{j,k=1}^3 {\epsilon_{ijk} \over 2} \bigg[ -\frac{c^i _{\hphantom{i}jk}}{2}\gamma_i^2+\frac{1}{2}\left(c^j _{\hphantom{j}ki} \gamma_j^2 +c^k _{\hphantom{k}ij} \gamma_k^2\right)\bigg]+\lambda_i \gamma_j \gamma_k
\label{eqSelfDual}
\eea
with a similar convention for the dot. In the last two equations, there is no summation in the last term, where $i,j,k$ are a cyclic permutation of $1, 2, 3$.

As already discussed, each of the above equations has two branches. From the self-dual four-dimensional side, this is determined by each of the two non-equivalent homomorphisms of $\mathfrak{g}\to so(3)$. From the three-dimensional viewpoint, this corresponds to the two non-equivalent flat connections of $so(3)$ over the group manifold $G$. This holds over the real numbers for I, II, VII$_0$ and IX, whereas VI$_{-1}$ and VIII require to pass to the complex. In all Bianchi A, the advertised correspondence holds as one-to-one for
 each class and each branch. It goes as follows:
\begin{enumerate}
\item In the cases  I, II, VII$_0$, IX, we must set for the metric  $\tilde\gamma_i= \gamma_i,  \forall i$, whereas there is a fine structure for the gauge field\footnote{Whenever
$
\exists\ j\neq k\neq i \neq j \text{ such that } c^{i}_{\phantom{i}jk}c^k_{\phantom{k}ij}\neq 0
$,
that is to say for Bianchi VII$_0$ and IX, as well as  VI$_{-1}$ and VIII, the branches are ``crossed''.
}: $\tilde\lambda_i= \lambda_i  \forall i$ for I and II, and  $\tilde\lambda_i= 1-\lambda_i  \forall i$
for VII$_0$ and IX.
 \item For VI$_{-1}$ and VIII, the correspondence is summarized in Table 2. Note that trivial automorphisms alow to displace the different entry in each case. A similar comment holds for the
connections given for all branches in Table 3.
\end{enumerate}

\begin{table}
\centering
\begin{tabular}{|c|c|}
  \hline
  Type & $\{ \tilde \gamma_1, \tilde \gamma_2, \tilde \gamma_3 \}$ \\ \hline \hline
  VI$_{-1}$ & $\{i \gamma_1, i \gamma_2, \gamma_3 \}$  \\ \hline
  VIII       & $\{\gamma_1, -i \gamma_2, -i \gamma_3 \}$ \\ \hline
\end{tabular}
\label{bt}
\caption{Mapping for metric coefficients in Bianchi VI$_{-1}$ and VIII.}
\end{table}
\begin{table}
\centering
\begin{tabular}{|c|c|c|c|c|}
  \hline
  Type & $\{ \lambda_1, \lambda_2, \lambda_3 \}_{(1)}$&$\{ \tilde \lambda_1, \tilde \lambda_2, \tilde \lambda_3 \}_{(1)}$&$\{ \lambda_1, \lambda_2, \lambda_3 \}_{(2)}$&$\{ \tilde \lambda_1, \tilde \lambda_2, \tilde \lambda_3 \}_{(2)}$ \\ \hline \hline
  VI$_{-1}$ & $\{0,0,0 \}$ &  $\{i,0,0 \}$  & $\{i,0,0 \}$  & $\{0,0,0 \}$   \\ \hline
  VIII       &  $\{0,0,0 \}$ &   $\{0,-i,-i \}$ & $\{0,-i,-i \}$  & $\{0,0,0 \}$  \\ \hline
\end{tabular}
\label{conne}
\caption{Mapping for connections in Bianchi VI$_{-1}$ and VIII, for each of the two branches.}
\end{table}

Although the classes  VI$_{-1}$  and  VIII are interesting neither for gravitational instantons nor for the Ricci flow (because of their complex nature) they are useful for setting the correspondence on a more universal ground. They might also play a more physical role in the search of self-dual solutions in a four-dimensional setting with signature $(-,-,++)$. We will not pursue this analysis here.

Besides the proper interest of the intriguing correspondence presented here, it has also the value of introducing a new kind of flow that deserves further investigation on its own right. In this line of thought, it should be mentioned that the analysis is usually considerably simplified by the integrable nature of the equations that appear in the framework of gravitational instantons. For Bianchi IX e.g., the Taub--NUT branch ($\tilde\lambda=0$ in Eqs. (\ref{ricIX})) is described by the Darboux--Halphen equations studied long ago \cite{Darboux, halph1, halph2} and rediscussed over the recent years both in the mathematical literature \cite{ Takhtajan:1992qb} or in more physical contexts \cite{ Cvetic:2001zx, Bakas:2006bz, Atiyah:1978wi}. The Eguchi--Hanson branch ($\tilde\lambda=1$ in Eqs. (\ref{ricIX})) leads to the Lagrange equations (special version of the Euler top equations) also solved long ago, possesing remarkable integrability properties  \cite{ Takhtajan:1992qb}. Similar results can be found for other Bianchi classes.
All these results available in the gravitational-instanton literature can be useful to tackle Ricci or related flows beyond the usual asymptotic analysis \cite{Isenberg:1992}.

At this stage of the developments the reader might feel frustrated by the two features of the flow (\ref{basic}, \ref{flat}), namely the absence of flow for $\tilde A$ and the flatness of the latter $so(3)$ connection. Their origin can be traced back as follows: $\tilde A$ originates from the anti-self-dual part of the four-dimensional Levi--Civita connection  which is required to have zero curvature $\mathcal{A}$ (Eq. (\ref{sdsol})).
Time-independence of the corresponding connection $A$   follows immediately and is translated on the three-dimensional side as $\dot{\tilde A}=0$.

More general flows with  non-vanishing $\dot{\tilde A}$ and $\tilde F $ could be reached if the self-duality condition on the \emph{Riemann} curvature were relaxed, and replaced by a milder condition that would still allow for a first-order description of time evolution without imposing the anti-self-dual Levi--Civita connection be a pure gauge. This is possible if we allow for a \emph{cosmological constant } in four dimensions. In this case, self-duality of the Riemann is traded for that of the Weyl tensor
\begin{equation}
\label{weyl}
 \mathcal{W}^{ab}=\mathcal{R}^{ab}-\frac{\Lambda_{\mathsf{c}}}{3}  \theta^a\wedge \theta^b
\end{equation}
and solutions of the Einstein equations ($R_{ab}=\Lambda_{\mathsf{c}} g_{ab}$) can thus be generated.
The anti-self-dual part of the connection now explicitly depends on time and the corresponding curvature is non-zero. This can be illustrated in the celebrated solution of Fubini--Study for Bianchi IX (describing in fact a metric on $\mathbb{C}P^2$). Translated in the three-dimensional side, the equation for the metric flow is still given by (\ref{basic}) but is now accompanied with a flow for $\tilde A$ and a constraint for $\tilde F$. The gauge field now carries a dynamics, which decouples when the cosmological constant is turned off.

As a conclusion of the above analysis we would like to make some final remarks. We should first stress the role of each of the ingredients that we have used throughout our developments. We worked in four dimensions where the orthogonal group is factorized into two three-dimensional subgroups and all degrees of freedom are reduced as self-dual plus anti-self-dual.  The foliation plus homogeneity assumption further introduce three-dimensional leaves and another three-dimensional group, $G$ related to $so(3)$ with non-trivial homomorphisms. Finally, the self-duality requirement effectively reduces the system to a three-dimensional one, whose dynamics turns out to be equivalent to a geometric flow on homogeneous three-manifolds endowed with an $so(3)$ gauge connection.

It is not clear to us whether the correspondence described here (involving in three dimensions the ``square-root'' of the four-dimensional metric) has a deep intrinsic geometrical meaning or is a rephrasing of the dynamics. The above arguments show however that this scheme is certainly not expected to generalize in higher dimensions and this should be opposed to another instance, already quoted, where a similar phenomenon occurs: the non-relativistic gravity discussed in \cite{Horava:2009uw}. There, the dimension is generic and the $D+1$ foliation is imposed at the level of the action, breaking \emph{explicitly} the diffeomorphism invariance. This drastically alters the structure of the propagating degrees of freedom, which in our case follow from a plain Einstein--Hilbert dynamics. The relation with a $D$-dimensional theory is set by the \emph{detailed-balance} condition, which resembles our self-duality condition and has the same effect, when combined with the foliation ansatz: the system effectively reduces to $D$-dimensions and the dynamics captured by a first-order flow equation.   This is valid for any $D$ because, as opposed to self-duality, the  \emph{detailed-balance} condition can be imposed at any dimension. The tensors which drive the geometric flow, however, depend on $D$. For $D=3$, e.g. these are the Ricci and Cotton tensors. This is another difference with our set-up, again bounded to the exclusive use of the Einstein--Hilbert action.


\section*{Acknowledgements}

The authors would like to thank C. Bachas, I. Bakas, M. Berthelson, P. Bieliavsky, M. Carfora, D. Friedan, G. Gibbons, J. Isemberg, D. L\"ust , A. Petkou, M.-T. Wang and E. Woolgar
for stimulating discussions. The 2008 Munich meeting \textsl{Field Theory and Geometric Flows}
as well as the 2009 GGI workshop \textsl{New Perspectives in String
Theory} and Pisa meeting \textsl{Geometric Flows in Mathematics and
Theoretical Physics} have been beneficial in many respects.  Marios Petropoulos would like to thank the Service de Physique
Th\'eorique de l'Universit\'e Libre de Bruxelles as well as the Service de M\'ecanique et Gravitation de l'Universit\'e de Mons--Hainaut for kind hospitality. Philippe Spindel thanks the IHES where the
present collaboration was initiated. John Estes acknowledges financial support from the Groupement d'Int\'er\^et Scientifique P2I. This research was partially supported by the French Agence Nationale pour la Recherche, contract  05-BLAN-0079-01.

\end{document}